\documentstyle [epsfig,emulateapj] {article}

\newcommand{\be}{\begin{eqnarray}}
\newcommand{\ee}{\end{eqnarray}}
\def\lsim{\mathrel{\rlap{\lower3pt\hbox{\hskip1pt$\sim$}}
    \raise1pt\hbox{$<$}}} %less than or approx. symbol
\def\gsim{\mathrel{\rlap{\lower3pt\hbox{\hskip1pt$\sim$}}
    \raise1pt\hbox{$>$}}} %greater than or approx. symbol
\newcommand{\msun}{\mbox{~$M_\odot$}}

\lefthead{Brown, Lee,\& Bethe} \righthead{Hypercritical Advection
Dominated Accretion Flow}

\begin{document}

%\leftline{Draft Version \today}

\title{Hypercritical Advection Dominated Accretion Flow}
\author{G. E. Brown, C.-H. Lee}
\affil{Department of Physics \& Astronomy,
        State University of New York,
        Stony Brook, New York 11794, USA}
\author{H.A. Bethe}
\affil{Floyd R. Newman Laboratory of Nuclear Studies,
       Cornell University, Ithaca, New York 14853, USA}

%---------------------------------------------------------------------

\begin{abstract}
In this note we study the accretion disc that arises in hypercritical
accretion of $\dot M\sim 10^8\ M_{\rm Edd}$ onto a neutron star while
it is in common envelope evolution with a massive companion. Such a
study was carried out by Chevalier (1996), who had earlier suggested that the
neutron star would go into a black hole in common envelope evolution.
In his later study he found that the accretion could possibly be held
up by angular momentum.

In order to raise the temperature high enough that the disc might
cool by neutrino emission, Chevalier found a small value of the
$\alpha$-parameter, where the kinematic coefficient of shear
viscosity is $\nu=\alpha c_s H$, with $c_s$ the velocity of sound
and $H$ the disc height; namely, $\alpha\sim 10^{-6}$ was
necessary for gas pressure to dominate. He also considered results
with higher values of $\alpha$, pointing out that radiation
pressure would then predominate. With these larger $\alpha$'s, the
temperatures of the accreting material are much lower, $\lsim
0.35$ MeV. The result is that neutrino cooling during the flow is
negligible, satisfying very well the advection dominating
conditions.

The low temperature of the accreting material means that it cannot
get rid of its energy rapidly by neutrino emission, so it piles
up, pushing its way through the accretion disc. An accretion shock
is formed, far beyond the neutron star, at a radius $\gsim 10^8$
cm, much as in the earlier spherically symmetric calculation, but
in rotation. Two-dimensional numerical simulation shows that an
accretion disc is reformed inside of the accretion shock, allowing
matter to accrete onto the neutron star with pressure high enough
so that neutrinos can carry off the energy.
\end{abstract}
\keywords{accretion, accretion disks -- binaries: close --
          stars: neutron -- black hole physics -- stars: evolution}

%---------------------------------------------------------------------
\section{Introduction}
\label{sec1}

In one stage of the theory of evolution of compact objects the neutron
star from the explosion of the primary progenitor enters into common
envelope evolution of the evolving secondary. Recently, developing
ideas suggested by Chevalier (1993) and later by Brown (1995),
Bethe \& Brown (1998) introduced hypercritical accretion into a
population synthesis of binary compact objects.
They found that in the standard scenario of compact binary evolution
the neutron star essentially always went into a low-mass black hole.

In scrutiny of hypercritical accretion during common envelope evolution,
Chevalier (1996) discussed the effects of rotation. He studied accretion
discs and, in particular, neutrino-cooled discs and advection dominated
discs. He chiefly considered discs with viscosity parameter
$\alpha \sim 10^{-6}$. Such small $\alpha$'s were needed
in order to reach high enough temperatures, in situations where gas
pressure predominates, so that the temperature was sufficient for neutrino
emission.

Calculations (Brandenburg et al. 1996; Torkelsson et al. 1996) suggest
that the higher values of $\alpha\sim 0.1$, which Shakura \& Sunyaev (1973)
suggested as arising from magnetic turbulence are more appropriate.
Chevalier (1996) notes that ``Radiation pressure
tends to dominate at higher values of $\alpha$; for $M=1.4~\msun$,
$\dot M =1~\msun$ yr$^{-1}$, and $r=10^8$ cm, the crossover point is
$\alpha\approx 10^{-6}$."

In this note we discuss the situation for $\alpha\sim 0.05$,
finding that radiation pressure predominates to the extent that we
can take the adiabatic index to be $\gamma=4/3$. We show the
resulting
hypercritical accretion to be essentially that of Houck \& Chevalier
(1991) and Brown (1995), although angular momentum was not
discussed in these earlier papers.

\section{Advection}

We consider super-Eddington accretion, so electromagnetic radiation
is negligible compared to the energy carried by the infalling matter.
We shall show that, for reasonable parameters, the temperature remains
sufficiently  low that neutrino emission is also small. Therefore the
accreting matter keeps nearly constant energy per unit mass.

The accreting matter orbits around the neutron star. Because of collisions
between the accreting particles, we may consider their orbits circular.
From Narayan \& Yi (1994), the orbital velocity is
   \be
   v^2= (r\Omega)^2\approx \frac 27 v_K^2
   \label{eq1}
   \ee
with $v_K=\sqrt{GM/r}$ is the Keplerian velocity and
$M$ the mass of the neutron star.

Because of viscosity, the velocity has a radial (inward) component
which is conventionally set equal to\footnote{
    Our $\alpha$ here is (2/3) of the $\alpha_{ss}$ of
    Shakura \& Sunyaev (1973), who took the kinematic coefficeint of
    shear viscosity to be $\nu=\frac 23 \alpha_{ss} c_s H$.
    Thus eq.~(\ref{eq2}) is that of Narayan \& Yi (1994),
    $\nu= \alpha c_s H$.
    Our treatment is equivalent to their advection dominated solution
    with their $f=\epsilon^\prime=\epsilon=1$.
    }
   \be
   v_r =\frac{dr}{dt}=- \frac 37 \alpha v_K.
   \label{eq2}
   \ee
We shall assume $\alpha=0.05$.

The potential energy per unit mass is $-GM/r$, and from eq.~(\ref{eq1})
the kinetic energy from the bulk motion of disc is $\sim \frac 12 v^2=GM/7r$.
Since we assume negligible heat loss, the heat energy in the local frame
rotating with the disc is
   \be
   \epsilon=\frac 34\times \frac 67 \frac{GM}{r},
   \label{eq3}
   \ee
where the factor 6/7 comes from the sum of the
local potential and kinetic energies in the disc, and
$1/3$ of $\epsilon$ having gone into $PV$ work (Enthalpy rather
than energy is conserved).
The density is related to the rate of energy accretion by
   \be
   \dot M=2\pi r \ 2 H\rho \ (-\dot r)
   \ee
where $H$ is the height of the accretion disc. An explicit solution of
the accretion problem, by Narayan \& Yi (1994) finds $H=\sqrt{\frac{2}{7}}
r\approx \frac 12 r$ which we adopt. Then, with eq.~(\ref{eq2}),
   \be
   \dot M=2\pi r^2 \left( \frac 37 \alpha v_K\right) \rho.
   \ee
Using eq.~(\ref{eq1}), and assuming $\alpha$ constant
   \be
   \rho\propto r^{-3/2}.
   \ee
The thermal energy per unit volume is
   \be
   \rho\epsilon =\frac{3 \dot M v_K}{4\pi\alpha r^2} \propto r^{-5/2},
   \label{eq7}
   \ee
which also gives the $r$-dependence of the pressure, $p=\rho\epsilon/3$.
Let us assume the mass of a neutron star to be $1.5\msun$ and its
radius $10$ km,\footnote{
     For the neutron star with $M = 1.5\msun$, the Schwarzschild radius
     is $R_{Sch}=4.4$ km, and the marginally stable orbit is
     $R_{MS}=3 R_{Sch}$, which is larger than the radius of neutron star.
     From Abramowicz et al. (1988), however, the inner disk can be
     extended to the marginally bound orbit $R_{MB}=2 R_{Sch}$ with
     hypercritical accretion.
     In this note, therefore, we assumed that the accretion disc extends
     to the surface of the neutron star.
     } then
   \be
   \frac{v_K}{4\pi r^2} = 1.2\times 10^{-3} {\rm cm^{-1}\ s^{-1}}.
   \ee
Let us further assume $\dot M=1\msun$ yr$^{-1}$ $=6.3\times 10^{25}$
g s$^{-1}$, and $\alpha=0.05$,
   \be
   \rho\epsilon =4.5\times 10^{24} {\rm erg\ cm^{-3}}.
   \ee
The energy density of electromagnetic radiation plus electrons and
positrons is
    \be
    \rho\epsilon = \frac{11}{3} a T^4
           = \frac{11}{4} (1.37) \times 10^{26} \ T_{\rm MeV}^4
             \ {\rm ergs\; cm^{-3}}.
%%%        &=& 3.8 \times 10^{26}\ T_{\rm MeV}^4
    \label{eq13}
    \ee
Equating the two expressions
    \be
    T=0.33 \ {\rm MeV}.
    \ee
This is only a moderate temperature. The emission of neutrino pairs,
according to Dicus (1972) and Brown \& Weingartner (1995) is
   \be
    \dot \epsilon_n &=& 1.0\times 10^{25} \left(\frac{T}{\rm MeV}\right)^9
    {\rm ergs\; s^{-1}\; cm^{-3}} \nonumber\\
    &=& 4.6 \times 10^{20} {\rm ergs\; s^{-1}\; cm^{-3}}.
    \label{eq6}
    \ee
This is the energy emission at the surface of the neutron star. It
can easily be seen that this quantity goes like $r^{-45/8}$. Multiplying
by the volume element $4\pi r^2 dr$ and integrating over $r$, we get
for the total energy loss by neutrinos.
    \be
    \dot W_n =\int\dot\epsilon_n 4\pi r^2 dr
    =2.2 \times 10^{39}\ {\rm ergs\  s^{-1}}.
    \ee
On the other hand, the energy advected by infalling matter is
    \be
    \dot M\epsilon &=& 6.3\times 10^{25}\ {\rm g \ s^{-1}}
    \times 1.3 \times 10^{20}\ {\rm erg\ g^{-1}} \nonumber\\
     &=&  8.2 \times 10^{45}\ {\rm erg\ s^{-1}}.
    \ee
So the fraction of energy radiated in neutrinos is
    \be
    F=\frac{\dot W_n}{\dot M\epsilon} = 2.7 \times 10^{-7}.
    \label{eq15}
    \ee
This proves our contention at the beginning of this section that the
energy lost during infall in neutrino emission is negligible.

If the rate of infall is changed or $\alpha$ is altered, the fraction
of neutrino radiation changes as
    \be
    F\sim \dot M^{5/4} \alpha^{-9/4}.
    \label{eq16}
    \ee
Only if $\alpha$ were reduced by a factor $10^{3.5}$, to about
$2\times 10^{-5}$, would neutrino radiation become appreciable.

%
%Once the accreted material is deposited on the neutron star,
%the density increases by orders of magnitude, and $\epsilon$ stays
%the same. The temperature increase substantially, so that nuclear burning
%(which goes all of the way to $^{56}Ni$ for $T\sim 0.35 $ MeV) takes
%place. The thermal energy can easily be radiated away by neutrinos.
%

Whereas the above might appear to solve the accretion problem, it
will not give a steady state solution because the accreted matter,
which is at too low a temperature to cool sufficiently rapidly by
neutrino radiation, cannot be fitted into the narrow gap between
neutron star and inner disc. Thus, it pushes outwards through the
accretion disc (Chevalier 1999). This was first seen in the
numerical calculation of Zel'dovich et al. 1972. In the
spherically symmetric case an accretion shock must be built into
the flow. Freely falling gas outside of this shock goes into
subsonic flow inside, with formation of an envelope at small radii
hot enough so that neutrinos can be emitted to carry off the
energy. Detailed calculation (Houck \& Chevalier 1991; Brown 1995)
gives
   \be
   r_{sh}\simeq 2.6\times 10^8 {\rm cm} \left(\frac{\dot M}{\msun\
   {\rm yr}^{-1}}\right)^{-0.37}
   \ee
Chevalier (1999) illustrates the necessity for the accretion shock
simply. Starting from hydrostatic equilibrium in the gap region
   \be
   \frac{dP}{dr}=-\rho\frac{GM\Gamma}{r^2}
   \ee
with $r$ initially set to $R$ because the gap between neutron star
and the innermost part of the accretion disc is narrow, and
$\Gamma=(1-R_{Sch}/r)^{-1}$, one can integrate from point A, the
inner edge of the HADAF, to point B, basically at the outer edge
of the neutron star. Taking
   \be
   P=K\rho^\gamma
   \ee
since the flow is adiabatic, and integrating from point A to point
B near the neutron star surface, one finds that
   \be
   \epsilon_B-\epsilon_A\approx\left[\frac{\gamma-1}{\gamma}\right]
   \frac{GM\Gamma}{R^2} (R_A-R_B)
   \ee
Now from eq.(\ref{eq3}) $\epsilon_A\sim GM/R$. For $\epsilon$ to
increase by a factor $\gsim 2$ so that neutrinos could carry off
the energy in going from point A to B would require
$(R_A-R_B)\gsim R$, violating the assumption that the gap is
narrow. This argument is general, not depending on the
dimensionality of the description.

The above discussion is at least roughly consistent with the
results of Armitage \& Livio (1999) who carry out 2D numerical
simulation of the hypercritical accretion. They find that an
accretion disc forms inside the accretion shock and with the
usually considered values of the viscosity the disc is probably
able to transfer mass inwards at the Bondi-Hoyle (hypercritical)
rate. Pressures at the neutron star are high enough so that the
accreted energy can be carried off by neutrinos. They warn,
however, that the pressure of a disc also makes the formation of
outflows or jets probable, which could reduce the accretion rates
onto the neutron star.

Our results up through eq.(\ref{eq16}) are consistent with those
of Minishige et al. (1997) for black hole disc accretion, once our
accretion rate is scaled down to the $\sim \dot M/\dot M_{Edd}\sim
10^6$ that they use. For a black hole accretion energy can be lost
from the system because the energy can be advected across the
horizon. However, even in the case of a black hole, an accretion
shock is developed in the numerical calculations of MacFadyen \&
Woosley (1999). Their rate of accretion is $\sim 0.1\msun\ {\rm
s}^{-1}$, $\sim 10^6$ times ours. In their evolution of the
accretion first the low angular momentum material in the equator
and material along the axes falls through the inner boundary onto
the black hole. An accretion shock at $\sim 350$ km with an
interior centrifugally supported disc then forms. The MacFadyen \&
Woosley (1999) calculation deals with a different situation from
ours; namely, advection of material that is originally He onto a
black hole. Here nuclear dissociation energy as well as neutrino
emission plays an important role in lowering the pressure. From
the Minishige et al. (1997) calculations at an accretion rate of
$\dot M\sim 10^6 \dot M_{Edd}$ into a black hole one can see the
formation of the accretion disc, but cannot discern an accretion
shock.

From the calculation most relevant to our work, that of Armitage
\& Livio (1999) one sees that the case the flow has initial
angular momentum, the main effect of the angular momentum in their
2D simulation is to set both accretion shock and accretion disc
into rotation. The matter then being advected smoothly onto the
neutron star. Thus, modulo the possibility of jets, the resulting
accretion is very much like that in the one-dimensional
calculations of Houck \& Chevalier (1991) and Brown (1995).

It should be noted, however, that in the one-dimensional case the
pressure drops off as $p\propto r^{-4}$, appropriate for a
$\Gamma=4/3$ accretion envelope (Chevalier 1989; Houck \&
Chevalier 1991; Brown \& Weingartner 1994), whereas in the ADAF
solution $p\propto r^{-5/2}$. The flatter profile in the latter
case makes it more difficult to reach the energy densities
necessary for neutrino emission.

Perhaps the most important result is that the orbital motion of the
accreting material is preserved. So the neutron star, in the natural course
of events, acquires an angular momentum,
    \be
    J=\Delta M \times r_N v_N
    \ee
where $\Delta M$ is the total mass accreted, $r_N$ is the radius of the
neutron star, and $v_N$ the corresponding orbital velocity,
    \be
    r_N v_N = \left(\frac 27 GM r_N\right)^{1/2}
            = 0.76 \times 10^{16}  \ {\rm cm^2\ s^{-1}}
    \ee
an appreciable angular momentum.

These consideration apply equally to accretion to a black hole.
The thermal energy, in this case, is not radiated away, but
$\epsilon_{th} c^{-2}$ is added to the mass of the black hole. If we
assume that the black hole has a mass $M_B=2\msun$, its
Schwarzschild radius is
    \be
    r_B=6 {\rm km}.
    \ee
From Abramowicz et al. (1988), for the hypercritical accretion,
the inner disk can be
extended to the marginally bound orbit $R_{MB}=2 R_{Sch}=12$ km.\footnote{
    Near the black hole, there should be relativistic corrections
    to the ADAF (Advection Dominated Accretion Flow) solutions.
    In this note, however, we assumed that the ADAF solution can be extended
    to the marginally stable orbit.
    }
So the orbital velocity at the inner disc boundary is
    \be
    v_B = 0.80 \times 10^{10} {\rm cm\ s^{-1}},
    \ee
and the specific angular momentum
    \be
    r_B v_B=  0.96 \times 10^{16} {\rm cm^2\ s^{-1}}
    \ee
about the same as for a neutron star. Going over to a Kerr Geometry
with decreased event horizon this will be lowered somewhat.
Bethe and Brown (1998) have considered a neutron star of $1.4\msun$
in common envelope with a giant, and have found that the neutron star
will accrete about $\Delta M=1\msun$ of the giant's mass and thereby
become a black hole. This accreted material will have an angular momentum
    \be
    J_B=\Delta M r_B v_B.
    \ee

Assuming that the core of the giant does not merge with the black hole,
this core will later become a supernova, yielding a second neutron star.
Gravitational waves will be emitted, and consequently later the two
compact objects will merge. It is important that the BH has a big
angular momentum $J_B$ which may be extracted
(Blandford \& Znajek 1977; Lee, Wijers, \& Brown 2000), and may
possibly give a gamma ray pulse: As we have seen, no special
mechanism is needed to give that angular momentum to the BH, but
it arises naturally as the black hole is formed.

%-----------------

\section{Discussion}

Although there is considerable uncertainty in the
$\alpha$-parameter in the viscosity, we believe our value to be in
a reasonable range. Torkelsson et al. (1996) have analyzed
simulations of the magnetic turbulence different groups used to
estimate the strength of the turbulent viscosity. Estimates of the
$\alpha$-parameter which range from 0.001 to 0.7 in the Shakura \&
Sunyaev (1973) definition, were verified using the same code for
all simulations. The higher values of $\alpha$ are the result of
an applied vertical (roughly equipartition) magnetic field.
Without this field the typical value is $\alpha\sim 0.005$.

Armitage \& Livio (1999) have raised the possibility of jets removing
the infalling material, so that it will not accrete onto the neutron star.
Jets are prevalent in situations with accretion discs, but the mechanisms
for driving them are poorly understood (Livio 1999).
Suggestions have been made that accretion discs could generate a
magnetically driven outflow, first by Blandford (1976) and
by Lovelace (1976).

If the magnetic energy above the disc is larger than the thermal and kinetic
energy densities, the ionized outflowing material is forced to follow field
lines, some of which form an angle with the perpendicular to the disc
surface. The material may then be accelerated by the centrifugal force like
a bead on a wire, as described by Livio (1999).

It would seem reasonable for jets to be driven in the range of accretion
rates $\dot M_{\rm Edd}$ to $10^4 \dot M_{\rm Edd}$, the
latter being the lower limit at which neutrinos can carry off the energy
sufficiently rapidly that hypercritical accretion can take place.
In this accretion interval, only $\dot M_{\rm Edd}$ can be accreted,
and since the surplus material must leave, it seems reasonable that magnetic
forces can collimate it in, at least, some cases. We believe that the jets
in Cyg X-3 may be an example.

It is more difficult to envisage such jet propulsion with hypercritical
accretion, especially at rates close to $\dot M\sim 10^9\ \dot M_{\rm Edd}$
where the Alfven radius comes right down to the surface of the neutron star;
i.e., where the ram pressure always exceeds the magnetic pressure,
unless the latter is increased by differential rotation, etc.
Of course, as in Eggum et al. (1988) who calculated only slightly
super Eddington accretion ($\dot M\sim 4\ \dot M_{\rm Edd}$)
a jet may be formed up the center of the disc axis, where there is little
infalling matter, but this would not drastically lower the accretion.

In fact, in a similar case of hypercritical accretion we have some
observational support for the main part of the matter being
accreted. In the evolution of the transient sources (Brown, Lee,
\& Bethe 1999) the H envelope of an $\sim 25\msun$ ZAMS is removed
in common envelope evolution with an $\lsim 1\msun$ main sequence
star. Following this, the Fe core of the massive star goes into a
black hole of mass $\sim 1.5\msun$ and most of the He envelope
must be accreted in order to obtain the final $\sim 7\msun$ black
hole. In general the He envelope will be rotating. In fact, in the
case of GRO J1655-40 arguments have been made (Zhang, Cui, \& Chen
1997) that the black hole is presently rotating close to the Kerr
limit although later studies (Sobczak et al. 1999; Gruzinov 1999)
lower the velocity to $\sim 0.6-0.7 c$.
Because the $\sim R_\odot$ radius of the He envelope is
$\sim 10^5$ times larger than the neutron star radius, even with
slow rotation of the envelope an accretion disc will be formed
before the matter can be advected into the black hole. The central
He density is $\rho\sim 10^5$ g cm$^{-3}$, about the same as the H
density in our work above in the inner disc, for $\dot M\sim
10^8\dot M_{\rm Edd}$. Thus, the difference is that the matter is
He, rather than H, in the transient sources. Clearly jet formation
does not strongly affect the amount of He accreted onto the black
hole.

It should also be realized that observed masses of neutron stars are not
very far below the Brown \& Bethe (1994) proposed upper mass
limit of $\sim 1.5\msun$.
Thus, only a small fraction of the Bethe \& Brown (1998) accretion of
$\sim 1\msun$ need be realized for the neutron star to go into a black
hole.

Since entropy gradients become negative, advection dominated flows
are unstable to convection, as found by Eggum et al. (1988) and
as worked out in considerable detail by Narayan \& Yi (1995).
Advection dominated accretion is found to dominate in a thick belt
about the equator, with somewhat reduced velocity, depending upon the
polar angle. Narayan \& Yi do find that jets can be driven in fairly
wide angles about the poles, but in agreement with Eggum et al. (1988) find
that very little mass is driven off in these jets.

Eq.~(\ref{eq7}) shows the energy density to go inversely with
$\alpha$, as $\alpha^{-1}$. Thus a much higher temperature is
reached for small $\alpha$'s, $\sim 10$ times greater for
$\alpha\sim 10^{-5}$ than for $\alpha\sim 0.1$ (Chevalier 1996),
the energy density going as $T^4$. The rate of cooling goes as
$T^9\propto \alpha^{-9/4}$ and increases a factor $\sim 10^9$ for
the small $\alpha$'s. In this regime of temperatures the disc is
cooled very rapidly by neutrino emission.

%---------------------------------------------------------------------
\acknowledgements

We would like to thank Roger A. Chevalier and Insu Yi for useful
suggestions and advice.
This work was partially supported by the U.S. Department of Energy under
Grant No.  DE--FG02--88ER40388.

%---------------------------------------------------------------------
\appendix
\section{Appendix: Advection Dominated Discs}

In this Appendix we give a more conventional derivation of
our results.

In the case that the disc is thin to neutrino emission, the
steady state disc equation can be written as (Shakura \& Sunyaev 1993;
Shapiro \& Teukolsky 1983, section 14.5)
   \be
   && \Sigma=2 H\rho,
   \label{Aeq1a}\\
   && \dot M=2\pi r\Sigma v_r={\rm const}
   \label{Aeq1b}\\
   && 6\pi r^2 H\alpha p =\dot M (GMr)^{1/2}
   \label{Aeq2}\\
   && H\dot\epsilon_{n}=\frac{3\dot M G M}{8\pi r^3}
   \label{Aeq3}\\
   && H=\left(\frac{p}{\rho}\right)^{1/2}\left(\frac{r^3}{GM}\right)^{1/2}
   \label{Aeq4}
   \ee
where we use the notation of Chevalier (1996).\footnote{
      In this note, we take the kinematic coefficient of shear viscosity
      $\nu=\alpha c H$, while  $\nu=\frac 23 \alpha c H$ is used
      in eq.~(17) of Chevalier (1996).
      %There is slip in eq.~(17) of compare with our eq.~(\ref{Aeq2}).
      }
Here $\Sigma$ is the surface density of the disc. Our equation of state is
   \be
   p=\frac{11}{4}\frac{aT^4}{3}
   \label{Aeq5}
   \ee
where we neglect the gas pressure. We assume that the neutrino pair
process (Dicus 1972) is the dominant source of neutrino cooling, using
the Brown \& Weingartner (1999) expression
    \be
    \dot \epsilon_n=1.0\times 10^{25} \left(\frac{T}{\rm MeV}\right)^9
    {\rm ergs\; s^{-1}\; cm^{-3}}.
    \label{Aeq6}
    \ee
Eq.~(\ref{Aeq3}) is however, not correct for the main part of our
accretion disc with radiation dominated  pressure,
because the temperature of the material is not high
enough for it to be cooled appreciably by neutrinos. In the absence of
appreciable cooling, we can use the Narayan \& Yi (1994) advection
dominated accretion, with self similar solution:
    \be
    v_r &\approx & -\frac{3\alpha}{5+2\epsilon}\left(\frac{GM}{r}\right)^{1/2},
    \label{Aeq7a}
    \\
    \Omega &\approx & \left(\frac{2\epsilon}{5+2\epsilon}\right)^{1/2}
    \left(\frac{GM}{r^3}\right)^{1/2}
    \label{Aeq7b}
    \\
    \rho &\approx &\frac{(5+2\epsilon)^{3/2}\dot M}
                {2^{5/2}3\pi\alpha r^{3/2} (GM)^{1/2}},
    \label{Aeq8a}
    \\
    p &\approx &\frac{2^{1/2}(5+2\epsilon)^{1/2}\dot M (GM)^{1/2}}
                {12 \pi\alpha r^{5/2}}
    \label{Aeq8b}
    \ee
where we take $\epsilon=1$ for the case of $\gamma=4/3$.

Energy is produced by viscous heating at a rate
    \be
    \dot \epsilon_{\rm prod}=
    \frac{3\dot M GM}{8\pi r^3 H} %\simeq 3.57\ \frac{\dot M GM}{(2r)^4}
    \label{Aeq9}
    \ee
where we use $H=\sqrt{\frac 27} r$ from eqs.~(\ref{Aeq4},\ref{Aeq8a},\ref{Aeq8b}).
In order to use eq.~(\ref{Aeq6}) we must obtain a temperature.
For $\alpha=0.05$, $M=1.5 \msun$, and
$\dot M=1 \msun$ yr$^{-1}=0.63\times 10^{26}$ g s$^{-1}$ we find
     \be
     \dot\epsilon_{\rm prod}
     =0.28\times 10^{20}\ r_8^{-4}\ {\rm ergs\; cm^{-3}\; s^{-1}}.
     \label{Aeq11}
     \ee
Using
     \be
     v_r =\frac{dr}{dt}= -\frac 37\alpha\sqrt{\frac{GM}{r}}
     \label{Aeq10n}
     \ee
we can integrate this expression:
     \be
     \epsilon_{\rm prod}
     &=& \int_0^t \dot\epsilon_{\rm prod} dt
     = \int_\infty^r \frac{\dot\epsilon_{\rm prod}}{v_r} dr
     \nonumber\\
     &=& -\int_\infty^r dr \frac{3\dot M GM}
             {\frac 37\alpha\sqrt{GM}8\pi (H/r)} r^{-7/2}
     = \frac 25 \times \frac{r}{\frac 37\alpha\sqrt{\frac{GM}{r}}}
     \times \frac{3\dot M GM}{8\pi r^3 H}
     \nonumber\\
     &=& \frac 25 \ \frac{r}{|v_r|}\ \dot\epsilon_{\rm prod},
     \ee
where the last expression shows $\epsilon_{\rm prod}$ to be (2/5) of
the value that would be obtained from dimensional analysis. The
quantity $(2/5)r/|v_r|$ can be interpreted as a dynamical time $\tau_d$
     \be
     \tau_d=\frac 25\frac{r}{|v_r|}= 1.3\ r_8^{3/2}\ {\rm sec}
     \ee
and
     \be
     \epsilon_{\rm prod} = \tau_d\ \dot\epsilon_{\rm prod}
     =3.6\times 10^{19}\ r_8^{-5/2}\ {\rm ergs\; cm^{-3}}.
     \label{Aeq12}
     \ee
We obtain a temperature of the advected material by relating
$T$ to the thermal energy density
    \be
    \epsilon_{\rm th} &=& a T^4
           = \frac{11}{4} (1.37) \times 10^{26} \ T_{\rm MeV}^4 \nonumber\\
        &=& 3.8 \times 10^{26}\ T_{\rm MeV}^4 \ {\rm ergs\; cm^{-3}}.
    \label{Aeq13}
    \ee
From eq.~(\ref{Aeq12}) we find \footnote{
   Eq.~(\ref{Aeq14}) shows that $T$ reaches $\sim 0.31$ MeV at the
   inner disc just outside the neutron star surface $10^6$ cm. There the
   dynamical time is only $\sim 1.3\times 10^{-3} $ sec, so the neutrino
   cooling by eq.~(\ref{Aeq6}) is negligible.
   }
   \be
   T_{\rm MeV}^4 =  0.96\times 10^{-7}\ r_8^{-5/2}
   \label{Aeq14}
   \ee
showing that $T^4$ goes as $r_8^{-5/2}$. The total cooling
by eq.~(\ref{Aeq6}) will be
   \be
   \delta\epsilon = \tau_d \ \dot\epsilon_n
   \label{Aeq15}
   \ee
so that
   \be
   \frac{\delta\epsilon}{\epsilon_{\rm prod}}
   =\frac{\dot\epsilon_n}{\dot\epsilon_{\rm prod}}
   =\frac{0.29\times 10^{-9}}{r_8^{13/8}}
   \label{Aeq16}
   \ee
Thus $\delta\epsilon/\epsilon_{\rm prod} \sim 1.6 \times 10^{-7}$
for $r=10^6$ cm at the inner disc just outside the neutron star,
in rough agreement with eq.~(\ref{eq15}).
In other words, temperatures of the advected material are so low that
the neutrino cooling is negligible. Thus, the material will be
deposited essentially adiabatically onto the neutron star, heating it.

%---------------------------------------------------------------------


\begin{thebibliography}{}
\bibitem[a]{Abramow88}
    Abramowicz, M.A., Czerny, B., Lasota, J.P., \& Szuszkiewicz, E. 1988,
    ApJ, 332, 646
\bibitem[a]{al99}
    Armitage, P.J., \& Livio, M., astro-ph/9906028
\bibitem[a]{bb98}
    Bethe, H. A., \& Brown, G. E. 1998, ApJ, 506, 780
\bibitem[a]{bm82}
    Begelman, M.C., \& Meier, D.L., ApJ, 253, 873
\bibitem[a]{blandford76}
    Blandford, R. 1976, MNRAS, 176, 465
\bibitem[a]{BZ77}
    Blandford, R.D., \& Znajek, R.L. 1977, MNRAS, 179, 433
\bibitem[a]{Brandenburg96}
    Brandenburg, A.,  Nordlund, \AA.,  Stein, R.F., \& Torkelsson U. 1996,
    ApJ, 458, L45
\bibitem[a]{brown95}
    Brown, G. E. 1995, ApJ, 440, 270
\bibitem[a]{brownbethe94}
    Brown, G. E., \& Bethe, H.A. 1994, ApJ, 423, 659
\bibitem[a]{BLB99}
    Brown, G.E., Lee, C.-H., Bethe, H.A. 1999, New Astronomy, 4, 313
\bibitem[a]{BLWB00}
    Brown, G.E., Lee, C.-H., Wijers, R.A.M.J., \& Bethe, H.A. 2000,
    Phys. Rep., to be printed
\bibitem[a]{brown97}
    Brown, G.E., \& Weingartner, J.C. 1994, ApJ, 423, 843
\bibitem[a]{Chevalier89}
    Chevalier, R.A. 1989, ApJ, 346, 847
\bibitem[b]{Chevalier93}
    Chevalier, R.A. 1993, ApJ, 411, L33
\bibitem[b]{Chevalier96}
    Chevalier, R.A. 1996, ApJ, 459, 322
\bibitem[b]{roger99}
    Chevalier, R.A. 1999, private communication
\bibitem[b]{Dicus72}
    Dicus, D.A. 1972, Phys. Rev. D, 6, 941
\bibitem[c]{Eggum88}
    Eggum, G.E.,  Coroniti, F.V., \&  Katz, J.I. 1988,  ApJ, 330, 142
\bibitem{gruzinov}
   Gruzinov, A. 1999, ApJ, 517, L105
\bibitem[c]{HC}
    Houck, J.C., \& Chevalier, R.A. 1991, ApJ, 376, 234
\bibitem[c]{LWB}
    Lee, H.K., Wijers, R.A.M.J., \& Brown, G.E. 2000, Phys. Rep.,
    to be published, astro-ph/9906213
\bibitem[a]{liang88}
    Liang, E.P. 1988, ApJ, 334, 339
\bibitem[a]{livio99}
    Livio, M. 1999, Phys. Rep., 311, 225
\bibitem[a]{lovelace}
    Lovelace, R.V.E. 1976, Nature, 262, 649
\bibitem[a]{MNHNS97}
    Mineshige, S., Nomura, H., Hirose, M., Nomoto, K., \& Suzuki, T. 1997,
    ApJ, 489, 227
\bibitem[a]{NY94}
    Narayan, R., \& Yi, I. 1994, ApJ, 428, L13
\bibitem[a]{NY95}
    Narayan, R., \& Yi, I. 1995, ApJ, 444, 231
\bibitem[a]{SS73}
    Shakura, N.I., \& Sunyaev, R.A. 1973, A\&A, 24, 337
\bibitem[a]{ST}
    Shapiro, S.L., Teukolsky, S.A. 1983, {\it Black Holes, White Dwarfs,
    and Neutron Stars"}, A Wiley-Interscience Publication.
\bibitem[a]{sobczak}
   Sobczak, G. J., McClintock, J. E., Remillard, R. A., Bailyn, C. D.,
   and Orosz, J. A. 1999, ApJ, 520, 776
\bibitem[a]{TZ77}
    Thorne, K.S., \& Zytkow, A.N. 1977, ApJ, 212, 832
\bibitem[a]{tork}
    Torkelsson, U., Brandenburg, A., Nordlund, \AA ., \& Stein, R.F. 1996,
    Astro. Lett. \& Comm., 34, 383
\bibitem[a]{Zeldovich}
    Zel'dovich, Ya.B., Ivanova, L.N., \& Nadezhin, D.K. 1972,
    Soviet Astronomy 16, 209
\bibitem[a]{ZCC}
    Zhang, S.N., Cui, W., \& Chen, W 1997, ApJ, 482, L155
\end{thebibliography}
\end{document}